\documentstyle[aps,prl]{revtex}

\begin{document}
\input epsf.sty
\twocolumn[\hsize\textwidth\columnwidth\hsize\csname %
@twocolumnfalse\endcsname
\draft
\widetext

\title{Antiferromagnetic Order of the Ru and Gd in Superconducting RuSr$_{2}$GdCu$%
_{2}$O$_{8}$}
\author{J. W. Lynn$^{1}$, B. Keimer$^{2}$, C. Ulrich$^{1,2}$, C. Bernhard$^{2}$, and
J. L. Tallon$^{3}$}
\address{$^{1}$NIST Center for Neutron Research, National Institute of \\
Standards and Technology, Gaithersburg, Maryland 20899 USA\\
$^{2}$Max-Planck Institute f\"{u}r Festk\"{o}rperforschung,\\
Heisenbergstrasse 1, D-70569 Stuttgart, Germany\\
$^{3}$Industrial Research Ltd., P. O. Box 31310, Lower Hutt, New Zealand}
\maketitle

\begin{abstract}
Neutron diffraction has been used to study the magnetic order in RuSr$_{2}$%
GdCu$_{2}$O$_{8}$. The Ru moments order antiferromagnetically at $%
T_{N}=136(2)K$, coincident with the previously reported onset of
ferromagnetism. Neighboring spins are antiparallel in all three directions,
with a low T moment of 1.18(6) $\mu _{B}$ along the {\em c} axis. Our
measurements put an upper limit of $\sim $0.1 $\mu _{B}$ to any net
zero-field moment, with fields exceeding $0.4T$ needed to induce a
measurable magnetization. The Gd ions order independently at $T_{N}=2.50(2)K$
with the same spin configuration.
\end{abstract}

\phantom{.}
]
\narrowtext

The ruthenate class of materials has been the focus of considerable work
recently because of their interesting magnetic and superconducting
properties. SrRuO$_{3}$, for example, is a {\em 4d} band ferromagnet that
orders at 165 K,\cite{SrRuO} while Sr$_{2}$RuO$_{4}$ is an exotic {\em p}%
-wave superconductor ($T_{C}=1.5K$).\cite{Superconductor} Of particular
interest here is the recent report of ferromagnetic ordering of the Ru at
133 K in RuSr$_{2}$GdCu$_{2}$O$_{8}$, while bulk superconductivity is
established at lower temperatures as observed in susceptibility and specific
heat.\cite{FerroSup,FerroSup2} In these hybrid ruthenate-cuprate systems
both the Cu-O and Ru-O planes form very similar square-planar arrays, and
the coexistence of superconductivity and long range magnetic order at high
temperatures is intriguing.\cite{Felner} Previous
``magnetic-superconductors'' such as the Chevrel phases ($RMo_{6}S_{8}$, $R$
= rare earth ion),\cite{Fischer} borocarbides ($RNi_{2}B_{2}C$),\cite
{Lynnboron} and cuprates ($RBa_{2}Cu_{3}O_{7}$ and related materials)\cite
{GschneiderMaple} show rare earth ordering at low temperature ($\lesssim 10K$
), and almost all are antiferromagnets that do not couple strongly to the
superconductivity. The rare occurrence of ferromagnetism,\cite{Fischer} as
found in $ErRh_{4}B_{4}$, $HoMo_{6}S_{8}$, and $HoMo_{6}Se_{8}$, revealed
the strongly competitive nature of these two cooperative phenomena in the
form of long wavelength oscillatory magnetic states at low temperature ($%
\lesssim 1K$) and a ferromagnetic lock-in transition that quenches the
superconductivity.\cite{ErRhB,HoMoS,HoMoSe} It would then be quite
interesting if RuSr$_{2}$GdCu$_{2}$O$_{8}$ were a ferromagnetic
superconductor with such a high magnetic ordering temperature, as this would
suggest\cite{Theory} a superconducting order parameter of the
Fulde-Ferrell-Larkin-Ovchinnikov type\cite{Ferrell} that could exhibit $\pi $%
-phase behavior.\cite{Buzdin} Our diffraction results, however, demonstrate
that the magnetic order of the Ru is predominantly antiferromagnetic,
instead making this by far the highest known antiferromagnetic ordering to
coexist with superconductivity. An upper limit of $\sim 0.1\mu _{B}$ is
obtained for the ferromagnetic component, consistent with recent
magnetization data, and the system is then similar to $ErNi_{2}B_{2}C$,
where a net magnetization develops below 2.3 K.\cite{FerroErNB} The Gd
moments also order magnetically, but at low temperatures in a manner
analogous to previous magnetic-superconductor systems.

A polycrystalline sample of RuSr$_{2}$GdCu$_{2}$O$_{8}$ was prepared by the
solid state reaction technique, using the $^{160}$Gd isotope to avoid the
huge nuclear absorption cross section of natural Gd. The single-phase sample
weighed $\sim 1.5$ g, has an onset superconducting temperature of 35 K and
bulk superconducting T$_{C}$ of 21 K, and is the identical sample used in a
previous susceptibility and neutron crystallographic study.\cite
{Crystallography} All the present neutron data were collected at NIST. BT-2
was employed at a neutron wavelength of 2.359 \AA , with a pyrolytic
graphite filter to suppress higher-order wavelengths. Polarized neutron
measurements were carried out with a Heusler monochromator and analyzer. A $%
^{3}$He refrigerator was employed for the lowest temperature measurements,
and a vertical field 7T superconducting magnet for the field measurements.
Small angle neutron scattering data were also collected with a wavelength of
5\AA\ on the NG-1 spectrometer from 6K to 300K. Statistical uncertainties
quoted in this article represent one standard deviation.

Fig. 1 shows a portion of the diffraction pattern obtained at a temperature
of 16 K. The peak at 23.5${^{\circ }}$ is the weak \{002\} nuclear Bragg
peak; for comparison, the strong \{103\}+\{110\} Bragg peak at 51.5${^{\circ
}}$ has 4809 counts/min. The peaks at 25.8${^{\circ }}$ and 30.8${^{\circ }}$
can be indexed as the $\{\frac{1}{2},\frac{1}{2},\frac{1}{2}\}$ and $\{\frac{%
1}{2},\frac{1}{2},\frac{3}{2}\}$ reflections. At a temperature of 150 K we
see that these peaks have completely disappeared, indicating that they are
magnetic and originate from the magnetic ordering of the Ru. There is no
change in the nuclear Bragg intensity, where a ferromagnetic component would
appear, as clearly indicated by the difference scattering\cite{Subtraction}
shown at the bottom of the figure. The temperature dependence of the
integrated intensity for the $\{\frac{1}{2},\frac{1}{2},\frac{1}{2}\}$ peak
is shown in Fig. 2. The solid curve is a simple mean field fit to estimate a
N\'{e}el temperature of 136(2) K. This is in excellent agreement with the
reported Ru magnetic ordering temperature.\cite{FerroSup}
\begin{figure} [t]
\vskip -5mm
\hspace{-5mm}
\centerline{\epsfxsize=3.6in\epsfbox{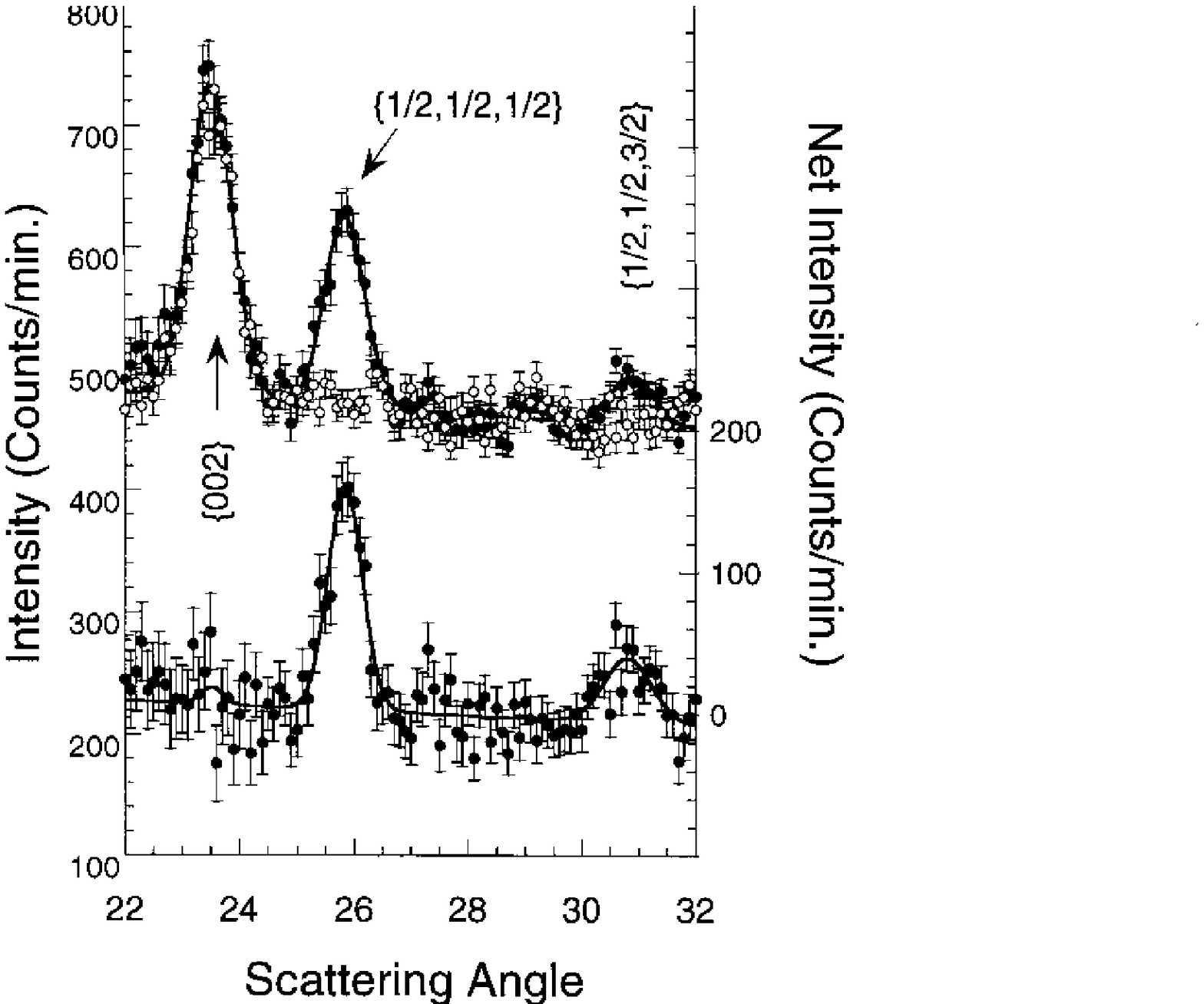}}
\vskip 2mm
\caption{Portion of a diffraction pattern at 16K, showing the weak nuclear \{002\} powder peak along with the $\{\frac{1}{2},\frac{1}{2},\frac{1}{2}\}$and $\{\frac{1}{2},\frac{1}{2},\frac{3}{2}\}$ Ru antiferromagnetic Bragg peaks. At 150 K the two magnetic peaks have disappeared, while there is no change in the \{002\} peak, where a ferromagnetic component would be observed. The difference scattering obtained by subtracting the two data sets is shown at the bottom.}
\label{Fig_1}
\end{figure}

The data in Fig. 1 show that the Ru moments order antiferromagnetically,
with nearest neighbor spins in all three crystallographic directions aligned
antiparallel. The magnetic scattering for a collinear structure is\cite
{Bacon} 
\begin{equation}
I_{M}=C\left| F_{M}\right| ^{2}\frac{m_{hkl}A_{hkl}}{\sin (\theta )\sin
(2\theta )}\left\langle 1-\left( \widehat{\tau }\cdot \widehat{M}\right)
^{2}\right\rangle
\end{equation}
where $C$ is an instrumental constant, $m_{hkl}$ is the multiplicity of the
powder peak with Miller indices $hkl$ for the reciprocal lattice vector $%
{\bf \tau }$, $A_{hkl}$ is the absorption factor, $\widehat{M}$ is a unit
vector in the direction of the moment, and the brackets indicate a
powder/domain average. The magnetic structure factor is given by 
\begin{equation}
F_{M}=\sum_{j=1}^{N}\left\langle \mu _{j}^{z}\right\rangle f_{j}({\bf \tau }%
)e^{i{\bf \tau \cdot r}_{j}}e^{-W_{j}}
\end{equation}
where $\left\langle \mu _{j}^{z}\right\rangle $ is the ordered moment and $%
f_{j}(hkl)$ is the magnetic form factor for the $j^{th}$ ion at position $%
{\bf r}_{j}$ in the unit cell, W$_{j}$ is the Debye-Waller factor, and the
sum is over all atoms in the unit cell. The magnetic intensities can be put
on an absolute scale by comparison with the nuclear intensities. The
intensity for the $\{\frac{1}{2},\frac{1}{2},\frac{3}{2}\}$ peak in Fig. 1
is clearly reduced in intensity compared to the $\{\frac{1}{2},\frac{1}{2},%
\frac{1}{2}\}$ peak, and the orientation factor in Eq. (1) then suggests
that the direction of the moment is along the tetragonal {\em c} axis. The
observed intensity ratio of the $\{\frac{1}{2},\frac{1}{2},\frac{1}{2}\}$ to 
$\{\frac{1}{2},\frac{1}{2},\frac{3}{2}\}$ peaks is 2.49(40), which is indeed
in good agreement with the calculated value of 2.21. Of course, with only
two observable magnetic peaks this moment direction assignment is tentative
rather than definitive. The ordered Ru moment at low temperatures is then
1.18(6) $\mu _{B}$, which agrees nicely with the moment of 1.05(5) $\mu _{B}$
obtained from susceptibility.\cite{FerroSup}
\begin{figure} [h]
\vskip -5mm
\hspace{-5mm}
\centerline{\epsfxsize=3.6in\epsfbox{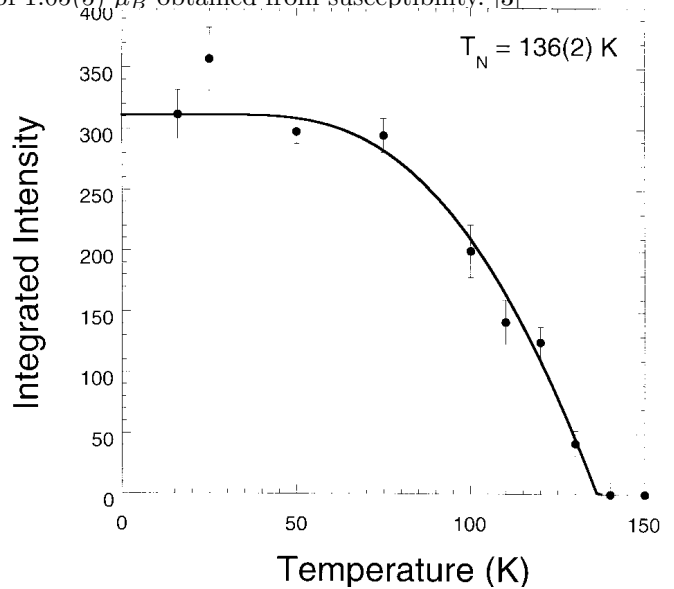}}
\vskip 2mm
\caption{ Integrated intensity of the $\{\frac{1}{2},\frac{1}{2},\frac{1}{2}\}$
Ru magnetic Bragg peak vs. T. The curve is a fit to mean-field theory.}
\label{Fig_2}
\end{figure}

Any ferromagnetic contribution to the scattering will occur at the same
positions as the nuclear Bragg peaks, and the data in Fig. 1 indicate no
magnetic contribution to the \{002\} peak within experimental error. We
measured the intensities of the ten lowest-angle nuclear Bragg reflections
above and below the Ru magnetic ordering temperature, and these data provide
an upper limit of $\sim $0.1 $\mu _{B}$ to any ferromagnetic component. This
result was substantiated by polarized beam measurements, although the error
limit was comparable to that obtained with unpolarized neutrons. We also
monitored the flipping ratio of the intensity transmitted through the sample
to determine if there were any depolarization of the beam as might be
expected for a ferromagnet. No change with temperature was observed.
Finally, we measured the small angle scattering on NG-1 in an attempt to see
any critical scattering associated with ferromagnetic correlations that
might develop, but no scattering was observed. Therefore the neutron data so
far do not reveal the ferromagnetic component associated with the Ru
ordering.

The field dependence of the magnetic scattering of the $\{\frac{1}{2},\frac{1%
}{2},\frac{1}{2}\}$ and $\{002\}$ peaks, corresponding to the
antiferromagnetic order and induced ferromagnetic moment, respectively, is
shown in Fig. 3. A temperature of 80 K was chosen for these measurements
since this is well below the Ru ordering temperature so that the sublattice
magnetization is near its saturated value, but it is high enough in
temperature that the Gd paramagnetic moment should not dominate the net
magnetization except at the highest fields. No significant change in either
intensity is observed up to $\sim 0.4T$. With further increase of field the
intensity of the antiferromagnetic reflection begins to decrease, while the
induced magnetization increases. At the highest field of $7T$ there is no
significant antiferromagnetic intensity remaining (as indicated by full
angular scans), while the induced magnetization corresponds to a net moment
of 1.4(1) $\mu _{B}$ perpendicular to {\em c}. The calculated induced Gd
paramagnetic moment is shown by the dashed curve, and for fields above $\sim
0.4T$ the data systematically lie above the curve indicating a Ru
contribution. However, the value is $\sim 0.2$ $\mu _{B}$, which suggests
the Ru moments are rotating into another antiferromagnetic (spin-flop)
structure, rather than becoming fully aligned with the field. Returning to
zero field, all the peaks recover their zero-field intensities, indicating
that the effect of the field is reversible, and also that no preferred
orientation of the loose powder particles occurred when the field was
applied. The Gd anisotropy is very small since it is an S-state ion. The low
spin-flop field and the lack of any field-induced preferred orientation then
suggests that the Ru crystalline anisotropy is also relatively weak,
indicating that a Heisenberg Hamiltonian is appropriate to describe the Ru
spin system.
\begin{figure} [b]
\vskip -5mm
\hspace{-5mm}
\centerline{\epsfxsize=3.6in\epsfbox{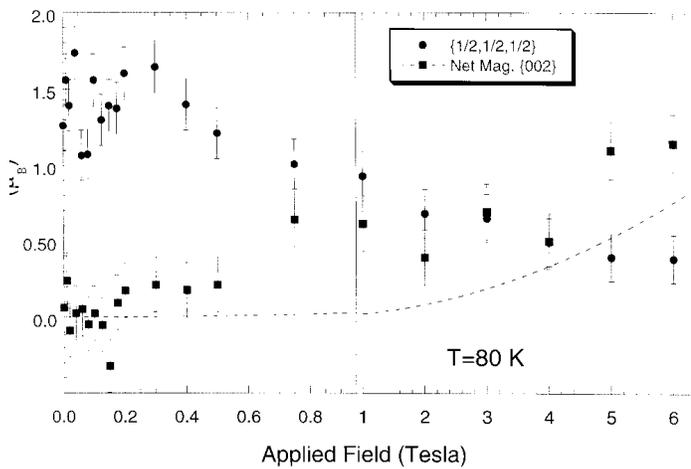}}
\vskip 2mm
\caption{Field dependence of the square of the antiferromagnetic moment and
induced ferromagnetic moment ($\propto $ normalized observed intensity) at
80 K. An expanded scale has been used at low fields for clarity. The dashed
curve represents the induced Gd moment.}
\label{Fig_3}
\end{figure}
\begin{figure} [h]
\vskip -5mm
\hspace{-5mm}
\centerline{\epsfxsize=3.6in\epsfbox{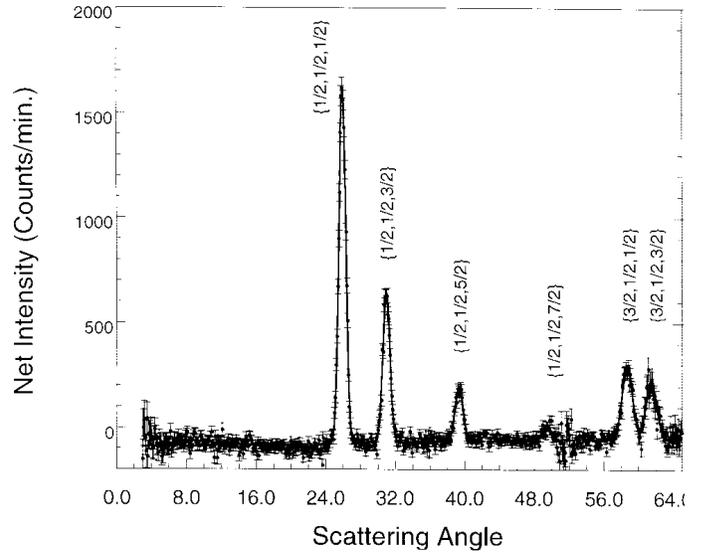}}
\vskip 2mm
\caption{ Magnetic diffraction pattern for Gd, obtained by subtracting the
data at 1.6 K from the data at 5K.}
\label{Fig_4}
\end{figure}

We now turn to measurements of the Gd magnetic order. Fig. 4 shows the
magnetic diffraction pattern, obtained by subtracting the data at 5 K from
the data at 1.6 K, below T$_{N}$.\cite{Subtraction} We see that the peak
positions and relative intensities are identical to that for the Ru, so that
nearest-neighbor Gd spins are also coupled antiferromagnetically along all
three crystallographic directions, with the moment direction along the
tetragonal {\em c }axis. The calculated and observed intensities then agree
to within the statistical uncertainties.

\begin{figure} [h]
\vskip -5mm
\hspace{-5mm}
\centerline{\epsfxsize=3.6in\epsfbox{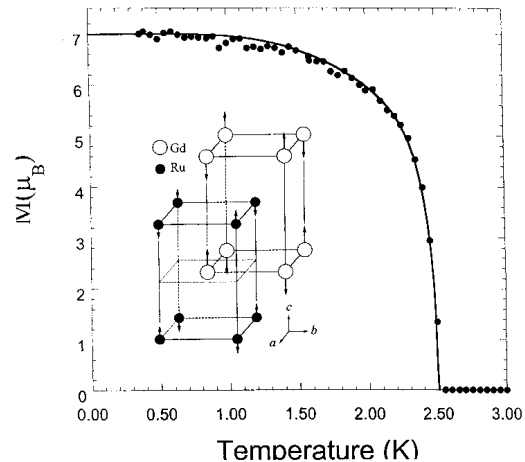}}
\vskip 2mm
\caption{Gd sublattice magnetization vs. T. The N\'{e}el temperature is 2.50
K. The inset shows the magnetic structures for the Ru and Gd.}
\label{Fig_5}
\end{figure}

The temperature dependence of the sublattice magnetization for the Gd is
shown in Fig. 5. We obtain the expected 7 $\mu _{B}$ moment within
experimental uncertainties at low T. Near T$_{N}$ a small correction has
been applied to the observed intensity to account for critical scattering,
and the solid curve is a fit to a modified power law, with a fitted ordering
temperature of 2.50(2) K. The sharpness of the ordering might at first be
surprising since the Ru and Gd magnetic structures are identical, and hence
one might expect them to be strongly coupled, smearing the Gd order
parameter. However, the Gd ions sit at the body-centered position of the
simple tetragonal Ru lattice. The antiferromagnetic magnetic structure for
the Ru then results in a cancellation of the average interaction between the
Gd and Ru (Fig. 4), rendering the two spin systems fully frustrated
(neglecting quantum fluctuations) with respect to each other and thus
behaving independently to a good approximation. It is noteworthy that this
frustration is relieved by a (zero-field) ferromagnetic component on the Ru
sublattice, so that the sharpness of the Gd order parameter is another
indication that the Ru magnetic structure can have only a modest net moment.

The Ru antiferromagnetic moment we observe is in good agreement with the
moment obtained from susceptibility, accounting for essentially the full
ordered moment. Our experimental upper limit of $\sim 0.1\mu _{B}$ on the
ferromagnetic component is consistent with the spontaneous moment derived
from low-field data, but is inconsistent with band structure calculations
predicting full ferromagnetic spin polarization of the Ru subsystem.\cite
{Theory} The crystallographic data\cite{Crystallography,FerroSup2} indicate
a rotation of the RuO$_{6}$ octahedra about the {\em c}-axis, with a small
rotation around an axis perpendicular to {\em c. }It is the latter rotation
that would be needed for conventional mechanisms such as antisymmetric
exchange or single-ion anisotropy to produce a canting and net Ru moment as
observed in a variety of measurements.\cite{FerroSup,FerroSup2} In any case,
the antiferromagnetism is dominant, and the coexistence of superconductivity
with the high ordering temperature and consequent large exchange
interactions of the Ru make this an especially interesting system for
further investigations.

Acknowledgments. We would like to thank R.W. Erwin and Q. Huang for their
assistance, and J. D. Jorgensen for communicating their results prior to
publication.

\end{document}